# Observation of Strong Coupling between an Inverse Bowtie Nano-Antenna and a Single J-aggregate

Adam Weissman[1], Maxim Sukharev[2,3] and Adi Salomon[1]*

[1] Department of Chemistry, Institute of Nanotechnology and Advanced Materials (BINA), Bar-Ilan University, Ramat-Gan 5290002, Israel

[2] College of Integrative Sciences and Arts, Arizona State University, Mesa, AZ 85212, United States

[3] Department of Physics, Arizona State University, Tempe, AZ 85287, United States

*E-mail: Adi.Salomon@biu.ac.il

**Abstract** - We demonstrate strong coupling between a single J-aggregate and an inverse bowtie plasmonic structure, when the J-aggregate is located at a specific axial distance from the metallic surface. Three hybrid modes are clearly observed, witnessing a strong interaction, with a Rabi splitting of up to 290 meV, the precise value of which significantly depends on the orientation of the J-aggregate with respect to the symmetry axis of the plasmonic structure. We repeated our experiments with a set of triangular hole arrays, showing consistent formation of three or more hybrid modes, in good agreement with numerical simulations.

# 1. INTRODUCTION

Strong interaction between photonic modes and molecular transition states leads to hybrid modes with new photo-physical and even chemical properties. [1–4] Therefore strong coupling has received considerable interest, both from chemistry and physics communities,[5,6] with applications to sensing,[7,8] nonlinear optics,[9] and photocatalysis,[10–12] as well as quantum information and communication technologies.[13–15]

In the strong-coupling regime, damping processes are small and the energy exchange between a molecular transition state and a photon state, occurs within a femto-second scale, faster than any other dissipative processes.[1,2] New hybrid states, separated by the Rabi splitting (RS) value, are formed. Those states, named polaritons, share both molecular and photonic characteristics,[16,17] Such states open new pathways for energy redistribution after molecular excitation and provide new routes for photochemistry on surfaces.[2,10,18]

Although plasmonic nano-structures possess ohmic losses, strong coupling can be observed at room temperature using plasmonic nano-structures, due to their tiny subwavelength mode volume, which can approach molecular-system dimensions.[1,19] Indeed, the strong- coupling regime has been observed in various plasmonic systems coupled to organic/inorganic transition states. [20–25] Coupling to molecular vibrational states has been shown too, leading to modification of a chemical bond frequency and, eventually, to chemical bonds alteration.[18] Strong coupling, hence, provides a new way to tailor the molecular surface potential energy, and it can lead to a modification of the very nature of the molecular system as well as to a shortening of natural processes/cycles in biological systems.[6,26–28]

Most of the experiments thus far have used an ensemble of emitters that were embedded irregularly in a thin polymer layer. Therefore, the observed strong-coupling signature was an average over the ensemble of emitters.[1,2]

The Baumberg group has performed a large set of experiments of strongly coupled systems made of oriented molecules and nanocavities at room temperature, where they used a gap-plasmon structure with oriented host-guest chemistry to form hybrid system with single molecules. Indeed, characteristic of mixing between light and matter were observed, with a Rabi splitting value of 90 meV for the case of single molecule coupling. In addition the Rabi splitting value was scaled linearly with the square root of the number of molecules (emitters).[29]

Interaction between dark modes of an individual plasmonic bowtie and few quantum emitters, has been shown as well by Bitton, O. et al.[30]

It is known that the axial distance of the emitter from the plasmonic surface plays an important role in determining the nature of its interaction with the surface plasmon (SP)[31–33] modes. If the emitter (molecule) is located far from the surface (> 100 nm), the interaction - if any - is very weak, due to the exponential decay of the plasmonic field.[16] On the other hand, if the emitter is in close proximity to the surface (~ 10 nm), the plasmonic field can affect not only the density of photonic states, but a strong interaction may occur, forming hybrid states as mentioned above.[34]

In the present work, we investigated the strong coupling regime between a single J-aggregate and an 'inverse' plasmonic bow-tie antenna milled in a silver thin film.[35–38] Deposited on a thin silica layer, the J-aggregate is located at a specific distance from the metal surface, typically at 15 nm. Moreover, the orientation of the J-aggregate rod with respect to the symmetry axis of the plasmonic structure can be precisely controlled.[39] Thus, our well-defined system provides a unique insight into the microscopic energy exchange mechanisms in such hybrid systems. Indeed, a strong coupling behavior with a clear splitting of 290 meV to a lower and an upper polaritons was observed. Moreover, in some cases a splitting to more than two hybrid states was observed. This may indicate ultra-strong coupling which leads to enhancement of inter-molecular interactions.[16,33,40] The nature of the coupling and the corresponding Rabi splitting depend both on the orientation of the J- aggregate rod with respect to the plasmonic structure and on the axial distance of the J-aggregate from the plasmonic surface.

Finally, we expand our study to the interaction of an about 5-nm thin J-aggregate layer with a periodic triangular plasmonic hole array. We also provide numerical simulations predicting the appearance of additional polaritons in the case of periodic hole arrays.

## 2. METHODS

**Sample preparation**

**Substrate cleaning** – Glass substrates (2cm X 1cm X 170μm) were immersed in a 70:30 mixture of sulfuric acid ($H_2SO_4$) and hydrogen peroxide ($H_2O_2$) solution at 90°c. The result was removal of all organic residues, and hydroxylation of the glass surface to OH groups.[41] That procedure was followed by an intensive rinsing with di ionized (DI, 18.6 MΩ) water and $N_2$ drying (99.999 % purity).

**Thin film deposition** - Sputtering physical vapor deposition (PVD, Q150T S , AVBA) was used to deposit Ag films with a controlled thickness of 220 nm, and ~2 nm roughness. The process was performed under medium vacuum (~$10^{-6}$ Bar).[42] E-beam evaporation was used to deposit 3 nm-30 nm transparent SiO2 layers on top of the metallic films, serving as a tunable nanometric spacer between the molecular system and the plasmonic system. The thickness of all films was measured by a Quartz Crystal Microbalance.

**Plasmonic structure fabrication** – A Gallium focused-ion beam (FIB, Helios NanoLabDualBeam 600, FEI) was used to mill plasmonic nano-cavities in the metallic thin films. To achieve an accurate nano-cavity shape, the fabrication process was calibrated for milling depth, beam size, and beam deflection. The integrated HR-SEM enabled the *in operando* nanometric imaging of the cavities. The thickness of the metallic film or multi-layered nano stratification, was validated using FIB cross section. The milled nano-cavities were triangular holes with a triangular side-length is 220 nm and a base of 200 nm.

**Formation of TPPS J-aggregate -** TPPS J-aggregates were prepared by dissolving Meso-tetra(4-sulfonatophenyl) porphyrin (0.1 to 10 mM) in DI water (5 ml), and adding 5-25 µL of 65% $HNO_3$ solution. The deposition procedure is reported elsewhere.[39]

**Spin coating** - Spin coating from anisole solution, was used to cover the plasmon-deposited-J- aggregates with a ~70 nm poly (methyl methacrylate) (PMMA) film and match the refractive index (RI) of all interfaces. Spin coating from water solution was used to coat the plasmonic structure directly with ~100nm of polyvinyl alcohol (PVA) in the control substrates. The polymer film thickness was measured using contact profilometry.

## Transmission spectra and hyperspectral Imaging

**Transmission spectra and imaging –** We used an Olympus inverted microscope (IX83 series) to measure the transmission spectra of the plasmonic structures. The samples were illuminated by collimated white light, in the bright-field mode. In cases of polarized transmission spectra, a rotatable linear polarizer was inserted in the optical path. The spectra and the hyperspectral images in Figures 2-7 and S1, S4, in the supporting information, were taken using 40x magnification (objective NA = 0.6), and were recorded using a spectrophotometer (IsoPlane SCT-320, Princeton Instruments) connected to a charge-coupled device camera (CCD, PIXS1024b, Princeton Instruments). Data analysis and imaging, including background subtraction, were done using MATLAB software.

**Cathodoluminescence measurements-** Hyperspectral images of the cathodoluminescence (CL) emission from the plasmonic sample (Figure 1d) were acquired on an Attolight Rosa 4634 CL microscope. The system comprised a tightly integrated achromatic reflective lens within the objective lens of a field-emission-gun scanning electron microscope (FEG-SEM). The focal plane of the light lens was matched to the FEG-SEM optimum working distance. A Czerny-Turner spectrometer was used to spectrally resolve the CL emission, which was finally sampled with a UV-Visible CCD camera. The electron beam had an acceleration voltage and emission current of 7 kV and 20 nA, respectively. CL spectra were acquired for each position of the electron beam on the sample, with 100 ms integration time. CL measurements were performed on silver-air interface (uncoated samples) and therefore the SP modes of the CL measurements are blue shifted by a factor of about 1.5 compared to the optical transmission spectrum. [43]

**Hybrid system micrographs** – the micrographs were taken on an upright Olympus microscope (BX51 series) with a dark-field aperture and captured on a complementary metal-oxide sensor color camera (Olympus CMOS SC100, 10.6 Mpix). [39] The alignment of the J-aggregate with respect to the plasmonic structure (inverse bow tie) was verified also by SEM and AFM microscopy (Bruker AXS, soft tapping mode).

**Simulations**

We numerically propagate Maxwell's equations using home-build codes and implement 3D-domain decomposition method to parallelize simulations. The material response is considered using conventional Drude-Lorentz model for silver with 5 Lorentz oscillators [45] Simulations are performed on 1152 processors with total number of grid points varying between $1.2 \times 10^8$ (for the period 400 nm) and $1.9 \times 10^8$ (for the period 500 nm). PML boundary conditions are implemented on top and bottom of the array. Linear transmission/reflection simulations typically take between 20 minutes to 40 minutes depending on the period of the array. Lastly, to simulate molecular response we implement rate equations approach,[46] and couple those with Maxwell's equations. The following set of parameters was used: transition dipole moment is 10 Debye, radiationless lifetime is 1 ps, pure dephasing time is 300 fs, and transition frequency is 2.53 eV.

## 3. RESULTS AND DISCUSSION

Our inverse bow-tie plasmonic antenna consists two triangular cavities of a side length of about 200 nm milled in 220 nm thin silver film. This plasmonic system was thoroughly studied before,[38,47] showing a spatial confinement of the electric field between the two cavities (Figure 1a), whose mode frequency is highly dependent on the distance between the holes. The plasmonic structure is covered by a thin silica layer of about 15 nm and a single J aggregate is deposited on top, followed by spin coating of a 70- nm PMMA layer (Figure 1b) in order to have the same refractive indices at both interfaces. The 15-nm thin spacer between the plasmonic structure and the J-aggregate prevents any direct chemical interaction between the two and maximizes the interaction through energy transfer process.[31–33] The TPPS J aggregate possesses two well defined transition states, S1 at 1.75 eV (706 nm) and S2 at 2.53 eV (489 nm),[39,48] potentially, and both of them can strongly interact with a given plasmonic system.[49] The absorption spectrum of a thin film J- aggregate is shown in Figure 1c, and mapping of the plasmonic system by cathodoluminescence is depicted in Figure 1d.[38] A confined mode at ~ 3.1eV nm is clearly observed between the cavities, at the interface with air.[38] Upon deposition of 70 nm of PMMA, the mode is red shifted to ~2.45eV,[43] and is thus resonant with S2.

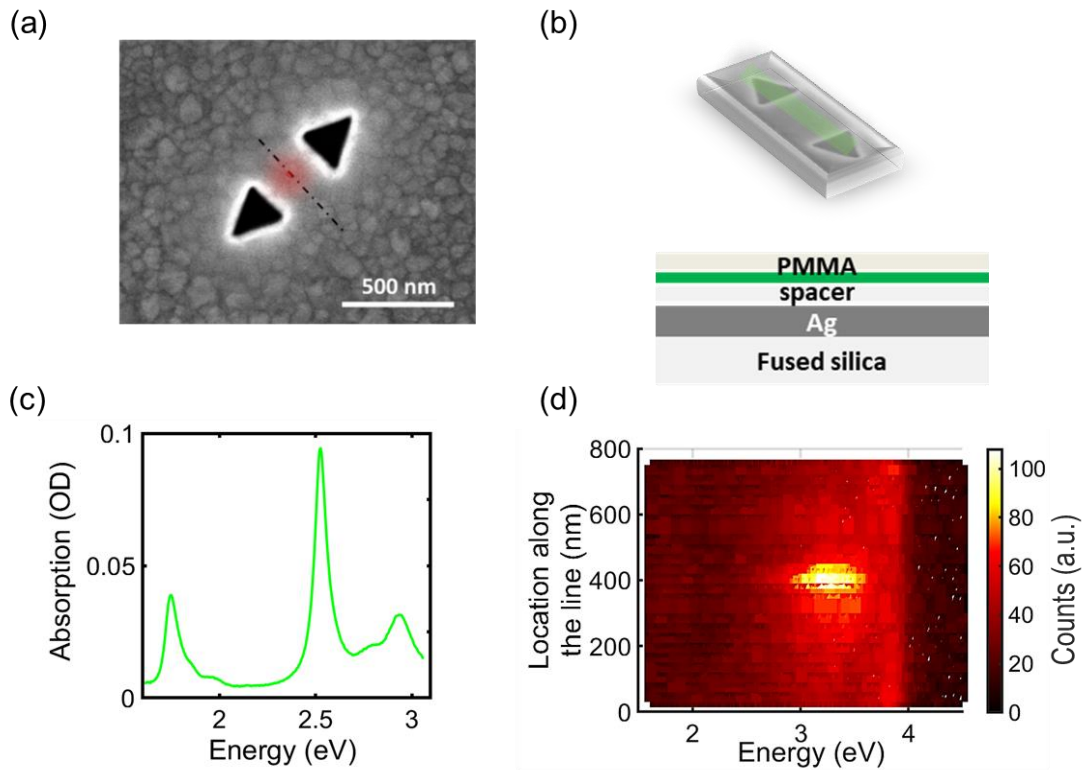

**Figure 1.** The inverse bowtie nano-antenna system, the molecular system, and the hybrid system. (a) HR-SEM image of the 'inverse' bow-tie plasmonic structure milled in a ~220 nm Ag film. The illustrated red dot indicates the mode confinement between the two cavities. (b) Illustration of a hybrid system and a its cross section, showing a 15 nm SiO2 spacer layer, 5 nm thin film of J aggregate and 70 nm of a capping layer made of PMMA. (c) Absorption spectrum of a J-aggregate thin film. The J-bands S1 (1.75eV) and S2 (2.53eV) are shown. (d) line scan from cathodoluminescence hyperspectral image (see figure 1a) showing a high energy mode confined onto the surface between the two cavities.

Figure 2a shows the transmission spectra of the plasmonic system, having a clear mode at about 2.45 eV (blue curve) together with that of the coupled system, with a single J-aggregate, deposited onto the plasmonic structure (green curve). Clearly, a hybridization between the two systems is observed, and three clear eigenmodes of the coupled system have emerged. We shall note that although interaction occurs between a single plasmonic mode and one J-aggregate, the signal-to-noise ratio of the raw data is very good (black line), and the coupling behavior is observed without further treatment of the results. For clarity we added the transmission spectra as a function of location, noting that the splitting occurs between the two cavities and does not spread over the entire structure, in contrast to the S1 spectral region (Figure 2b).

Such a behavior can be realized if the coupling strength between the two different modes is greater than the inverse lifetime of the molecular transition state and that of the localized mode. [2] It ensures that the coherence of the molecular transition state as well as the SP mode retains for a longer time than the Rabi oscillation period. The Rabi splitting value hence should be greater than the linewidth of both the SP mode and the molecular transition state (i.e., show a shorter life time). These conditions are met in our hybrid system and a broadening of the coupled system with respect to the plasmonic system is clearly seen. The observed spectrum of the coupled system is not a convolution of the two sub-systems. On the contrary, when the same molecular system is deposited onto the other side of the substrate (i.e., onto the glass), a dip in the spectrum is observed, which is attributed to a trivial filtering effect with no line broadening or an appearance of a third mode (Figure S1 in the supporting information).

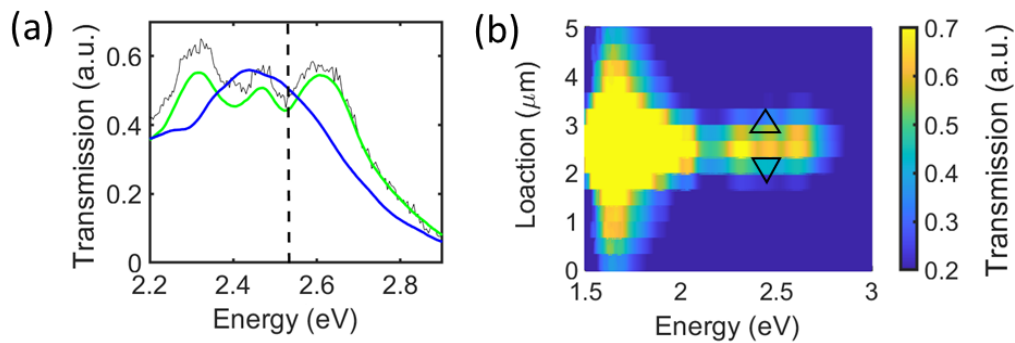

**Figure 2.** Transmission spectra and spectral imaging of the inverse bowtie nano-antenna - J-aggregate hybrid system. The dashed line at 2.53 eV indicates the J-aggregate transition state. The unsmoothed spectrum of the hybrid structure (black) is shown as well. (b) spectral imaging linescan across the coupled system, showing the S2 splitting confined at a region between the two cavities, while S1 enhancement is spread across the whole structure. Cavity distance base to base is 280 nm.

Next, we studied the nature of interaction with three distinct geometrical orientations of the J-aggregate with respect to the plasmonic interaction axis, parallel (0°), orthogonal (90°) or at 80° (see Figure 3 and Figure S2 in the supporting information). The transmission spectra have been taken at the same polarization state of the incoming field, i.e., parallel to the plasmonic interaction axis. A maximum Rabi splitting value of about 290 meV was observed when the orientation of the J-aggregate is parallel with respect to the plasmonic structure axis, yet, by and large the same optical signatures were observed in all three cases. The energy positions of the lower and upper polaritons are dependent on the J-aggregate orientation. Yet, the third mode, is pinned at the same energy position at about 2.47 eV, revealing its molecular nature (see table in Figure 3b). Therefore, it is plausible that this mode is related to enhanced interactions between the monomers in the aggregates, leading to an additional collective mode.[16]

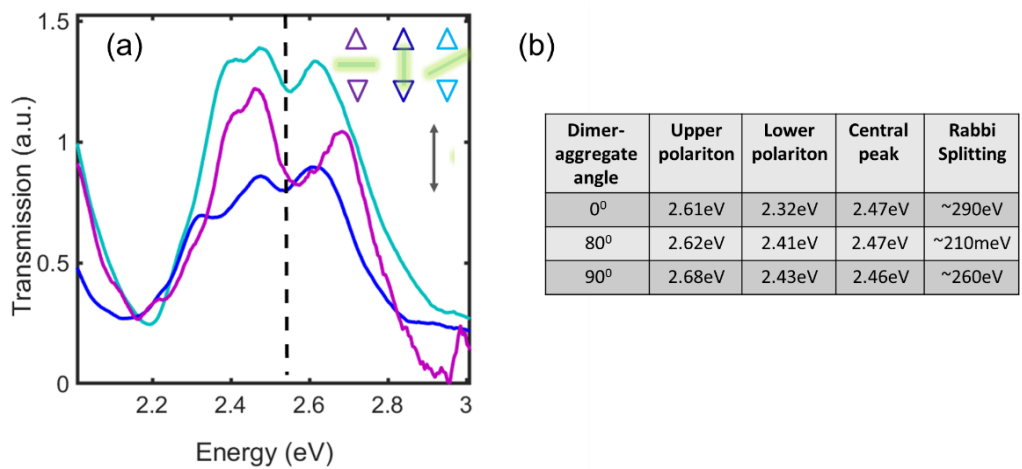

**Figure 3.** Orientation dependance of the inverse bowtie nano-antenna - J-aggregate hybrid. (a) J-aggregates deposited on the same plamsonic structure in three different orientations (inset). Upon strong interaction, a splliting to three hybrid modes is observed, with maximal splitting in parallel orientation. (b) a table which summerizes the results presented (a). Cavity distance base to base is 280 nm. The dashed line at 2.53 eV indicates the J- aggregate transition state.

The energy of the confined mode onto the flat surface between the two cavities, depends on the distance between them and hence, it can be systematically tuned along the visible range as was shown elsewhere.[38] Figure 4 shows a set of hybrid systems in which the plasmonic mode is on/off resonance with respect to the molecular transition state. First a dispersion of the polaritons is observed. Second, the collective mode which is clearly observed when resonant with the J-aggregate transition state, is not pronounced when the plasmonic mode is off resonant, in agreement with theoretical work.[16]

The strong interaction is very sensitive to the axial distance of the emitter (J-aggregate) with respect to the surface. Indeed, when the J-aggregate is deposited 30 nm above the surface no splitting is observed, assuring our experimental study (see dotted plot in Figure 4).

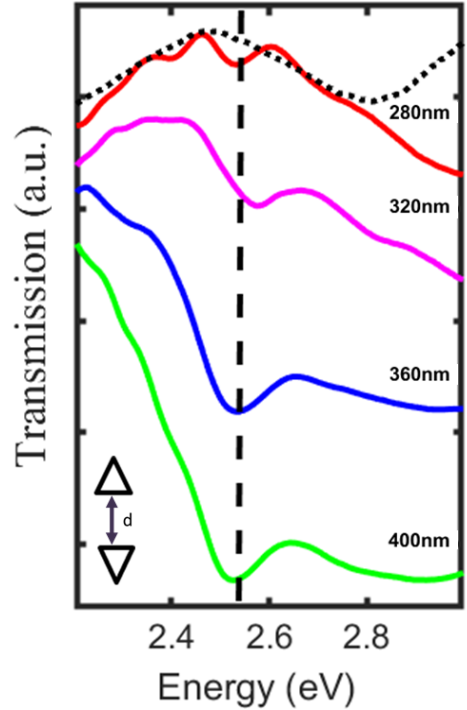

**Figure 4.** Hybrid modes as a function of the inter-cavity distance, *d*. The J-aggregate are deposited ~ 15 nm axial distance from the surface (see Figure 1b). The dashed line at 2.53 eV indicates the J- aggregate transition state. At resonance (280 nm base-to-base distance) splitting to three eigenmodes is observed. When the J -aggregates are deposited at 30 nm from the plasmonic system (black dotted plot) no splitting/interaction is observed.

To address the mechanism of the appearance of this collective mode, we extended our study to a set of periodic cavity arrays with identical sub-unit triangular cavities. A SEM micrograph of a typical hexagonal plasmonic array is depicted in Figure 5a, alongside with simulations of transmission through the arrays for four different periodicities (Figure 5b). The dielectric media refractive index at the two interfaces was fixed to be ~ 1.5, to mimic the reported experimental plasmonic system. Two plasmonic modes are observed, one at about molecular energy transition S1 and the other at about molecular transition S2 (for periodicity of 394 nm). Those peaks are red shifted with increasing periodicity as expected according to Bragg scattering.[50] The inset in Figure 5b shows a correlation between experiment and simulation. Calculation of the electric field distribution at 1.75 eV are and at 2.53 eV are shown in Figures 5c and 5d. At 1.75eV the hot spots are located at the triangular side length, whereas at 2.53 eV, they are localized between the triangular cavities onto the surface.

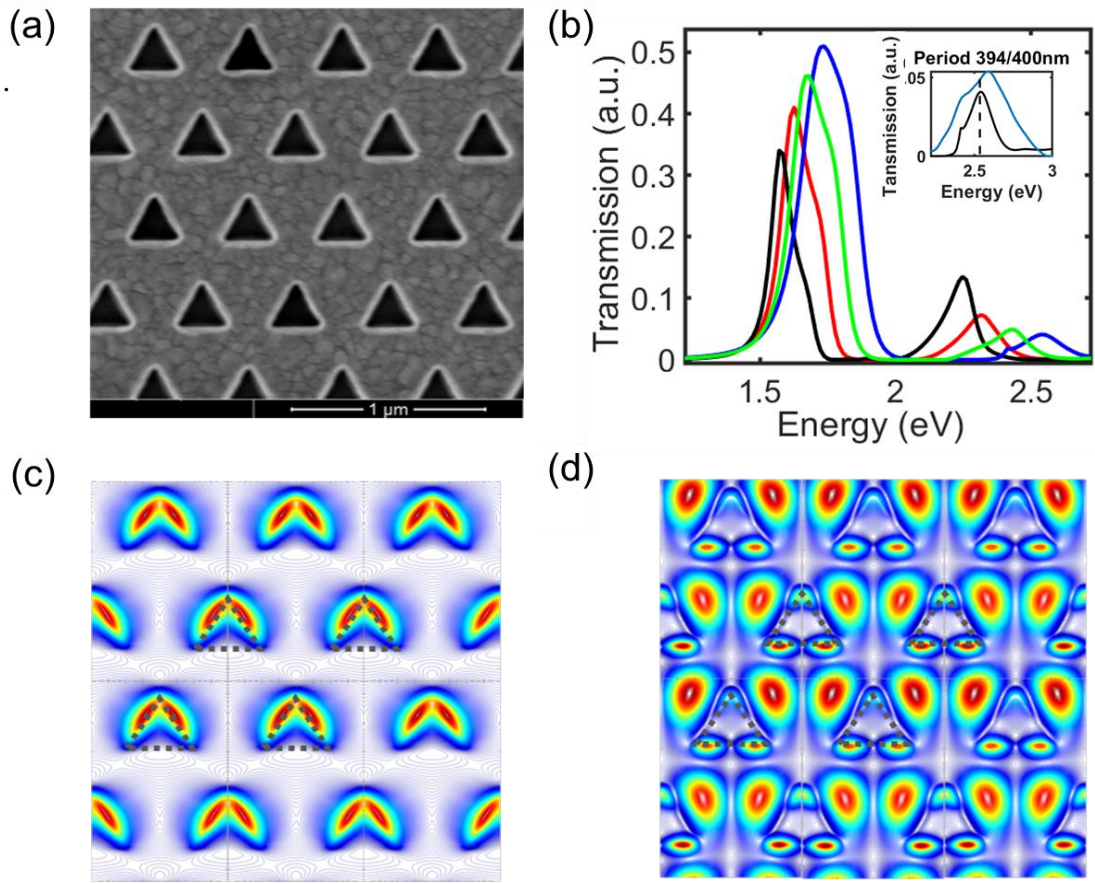

**Figure 5**: Plasmonic triangular-hole-array and its mode distribution. (a) SEM micrograph of the plasmonic structure (b) numerical simulation of transmission spectra of a set of Triangular-hole-array covered with 15 nm of SiOx. The array periodicity is: 394 nm (blue) ,430 nm (green), 466 nm (red), 502 nm (black). Inset - an agreement between experimental (blue) and simulation (black). The dashed line at 2.53 eV indicates the J- aggregate transition state. (c) and (d) show spatial distribution of the simulated electric field at 1.74 eV and at 2.54 eV respectively. The plasmonic modes at these two energies, reside either in the cavity walls (c), or in hot spots between the cavities (d).

The transmission spectra for the hybrid system are simulated for two molecular concentrations, i.e. $10^{26}$ m$^{-3}$ and $2\times10^{26}$ m$^{-3}$. In addition to the splitting of the modes, the plots in Figure 6a show that upon increasing the molecular concentration , a third collective mode is observed, an indication to intermolecular interaction of molecules immersed in the plasmonic field (see Figure S3 in the supporting information for more simulations).[16,21] An agreement between experimental (magenta) and simulation (blue) is depicted in Figure 6b, where in both cases, the coupling results as appearance of more than two eigenmodes (see Figure S4 in the supporting information).

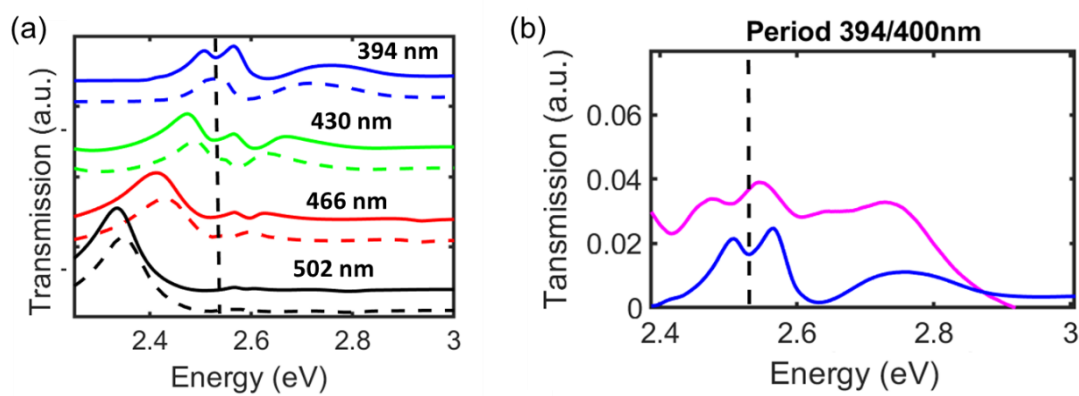

**Figure 6.** Concentration and spectral-tuning dependence in the Plasmonic triangular-hole-array – J-aggregate hybrid system. (a) simulation of transmission spectra of the system in varying pitch, for two molecular concentrations of the TPPS : $1 \times 10^{26}$ m$^{-3}$ , (dashed) and $2 \times 10^{26}$ m$^{-3}$ (solid line) showing the evolution of the collective mode with both resonance and concentration (pitch indicated in the inset). (b) Agreement between experimental (blue) and simulation (magenta) for molecular concentration of $2 \times 10^{26}$ m$^{-3}$ and pitch distance of ~400nm. Both theory and experiment support the appearing of the collective mode. Ag film thickness is 220 nm. The system is covered with 15 nm of SiO2 and the emitter is deposited ontop. The dashed line at 2.53 eV indicates the J- aggregate transition state.

Experimental results of detuning the plasmonic mode energy by +/- 100 meV with respect to $S_2$, are shown in Figure 7. The left panel in Figure 7a shows the transmission spectra of a set of triangular periodic hole arrays covered with 15 nm of SiO$_x$, and a clear dispersion of the plasmonic modes which is ranging from 2.6 eV to 2.4 eV is observed. Upon depositing of a single layer of J-aggregate TPPS a clear splitting in is observed (Figure 7a – right). In Consistency with the results presented above, more than two modes are observed for a small detuning. Could be that other molecular modes are enhanced due to the plasmonic field.[51]

The polaritons peak positions of those set of coupled systems are plotted in Figure 7b, together with the plasmonic modes of the solely triangular hole arrays (black dots). A typical signature of strong coupling, an anti-crossing behavior, is observed, where the original SPs modes (black dots), undergo a splitting at $S_2$ state of the J-aggregate. Consistently, a splitting to more than two modes have been observed. All the modes are plotted, and we cannot clearly identify their origin, since they may originate from enhanced interaction in the molecular system, or coupling to other plasmonic modes. [16,51,52]

The anti-crossing behavior is calculated where the lower and the higher polaritons, $E_l(k), E_u(k)$, are deduced according to equation (1) [53]

$$E_{u,l}(k) = \left\{[E_{pl}(k) + S_2]/2 \pm \sqrt{\Delta^2 + \left\{\frac{[E_{pl}(k) - S_2]^2}{4}\right\}}\right\} \qquad (1)$$

where, $k$ is the in-plane wave vector (for each periodicity ($k = \frac{2\pi}{P}$), and $2\Delta$ is the Rabi splitting value of 180 meV, representing the interaction strength. On resonance, the splitting is symmetric, and two eigenmodes with the same character are formed. However, at large detuning, the coupling reduces, the splitting is non-symmetric, and the two newly formed states have a different characteristic.[1] At the asymptotic line close to $S_2$, the hybrid states have a molecular-like character, whereas, close to the SPs line they have a plasmonic-like state. [16]

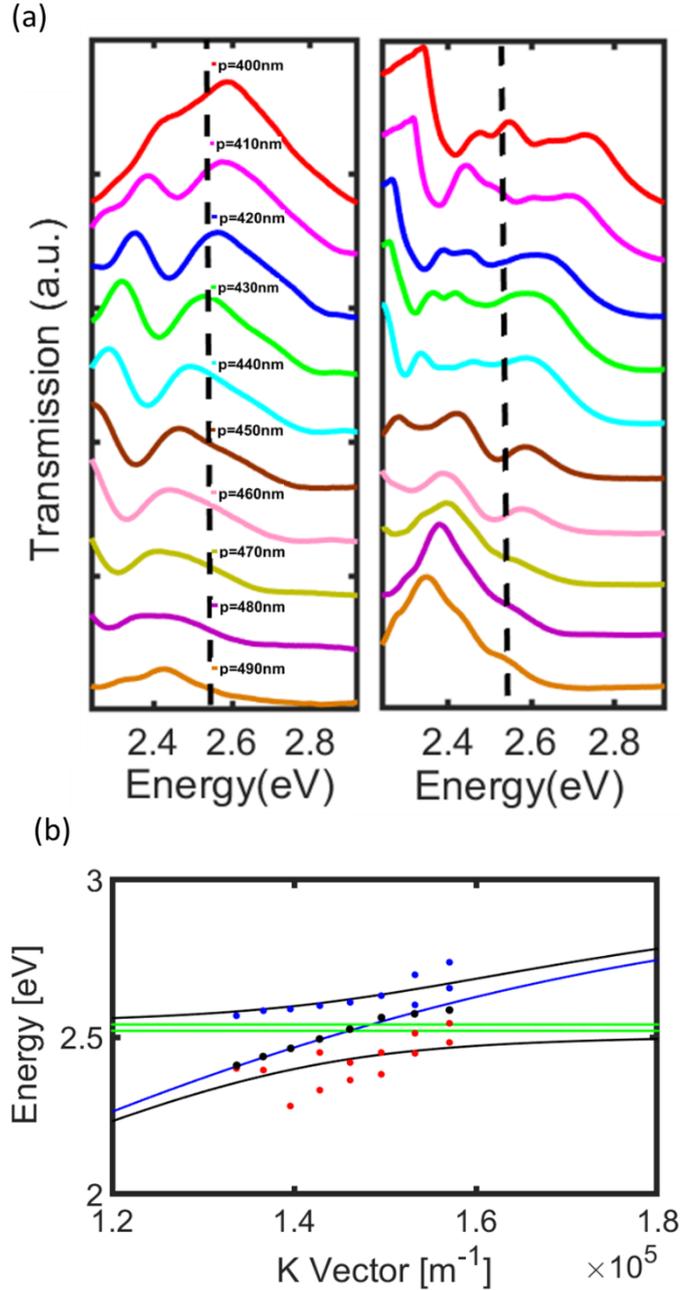

**Figure 7.** Anti-crossing behavior in the plasmonic triangular-hole-array J-aggregate hybrid system. (a) transmission spectra of a set of plasmonic Triangular-hole-array covered only with 15 nm of SiO2 (left) and a similar set in the hybrid system, on which a single J-aggregate layer is deposited (right). The periodicity (center-to-center) is indicated on the curves and apply to both panels. The dashed line at 2.53 eV denotes the transition states of **S₂** state. All spectra were measured under non-polarized light, at the same illumination conditions. (b)- analysis of (a), showing anti-crossing behavior of the hybrid system. The green nondispersive line indicates the J-aggregate transition state, the black dots are the extracted plasmonic modes from (a- left panel) for each periodicity ($K = \frac{2\pi}{P}$) and the blue/red stars are the extracted peaks from (a –right panel), indicating the upper/lower polaritons respectively. The black line is the calculated anti crossing behavior with RS value of 0.18 eV.

## 4. CONCLUSIONS

To conclude, we show a strong coupling behavior between a single plasmonic mode and a single J-aggregate, where an additional mode is observed in addition to two new polaritons. Although the coupling of the inverse bow-tie is to a single emitter, 290 meV Rabi splitting value is observed, which is about 12 % of the transition energy of the emitter, indication for ultra-strong coupling. The energy position of the new eigenmodes of the hybrid system are sensitive to the orientation of the J-aggregate with respect to the plasmonic structure interaction axis, whereas the additional mode, seems to be pinned at the same energy. At a distance of about 30 nm from the surface no splitting is observed, indicating that indeed strong interaction occurs in proximity to the plasmonic field. Our study was extended to periodical array system, in which a similar behavior was observed and was supported by numerical simulations. We expect such systems to enhance nonlinear effects such as second harmonic generation and surface enhanced Raman scattering.

## Author Contributions

The manuscript was written through contributions of all authors. All authors have given approval to the final version of the manuscript.

## ACKNOWLEDGMENTS

The study is supported by German-Isreali Foundation (Grant no. 203785) and is partially supported by the ISF-NSFC grant no. 2525/17

M.S. is grateful for the financial support provided by Air Force Office of Scientific Research under Grant No. FA9550-19-1-0009

# Supporting Information

# Observation of Strong Coupling between an Inverse Bowtie Nano-Antenna and a Single J-aggregate

Adam Weissman[1], Maxim Shukharev[2] and Adi Salomon[1]*

[1] Department of Chemistry, Institute of Nanotechnology and Advanced Materials (BINA), Bar-Ilan University, Ramat-Gan 5290002, Israel

[2] College of Integrative Sciences and Arts, Arizona State University, Mesa, Arizona 85212, United States

*E-mail: Adi.Salomon@biu.ac.il

.

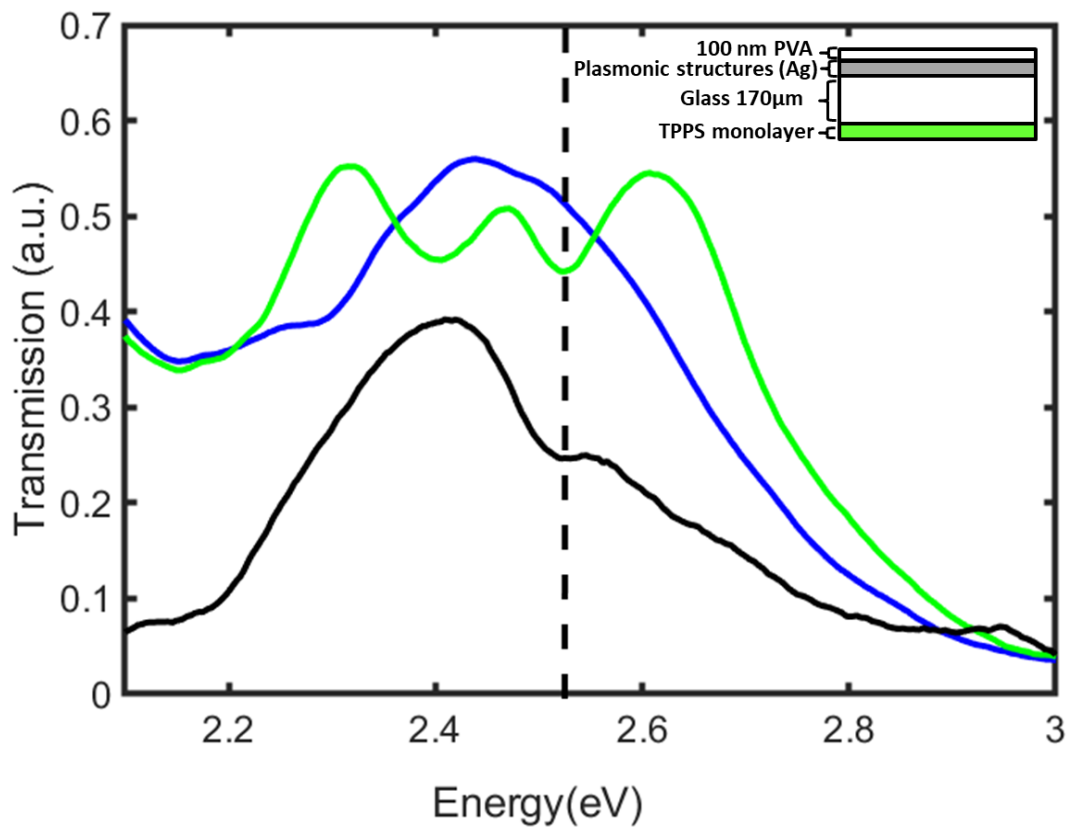

**Figure S1. Control experiment.** Transmission spectra of the plasmonic structure (blue) and the hybrid system (green) as in Figure 2, compared to a case in which an aggregate was deposited on the other side of the glass substrate (black), in the same spatial position, but in 170 microns axial distance from the plasmonic structure (see inset)

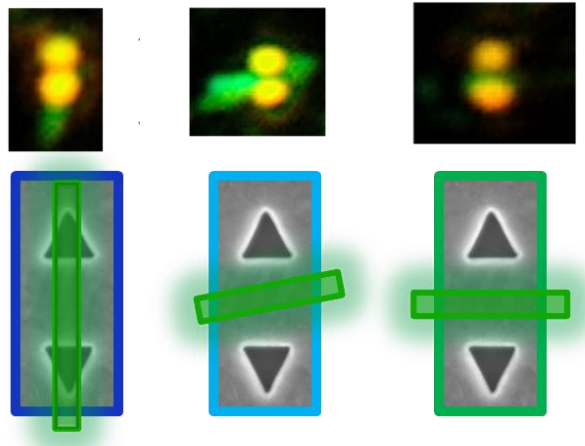

**Figure S2.** Dark-field micrographs (top), and corresponding illustrations (bottom) of the three different orientations of the hybrid structures.

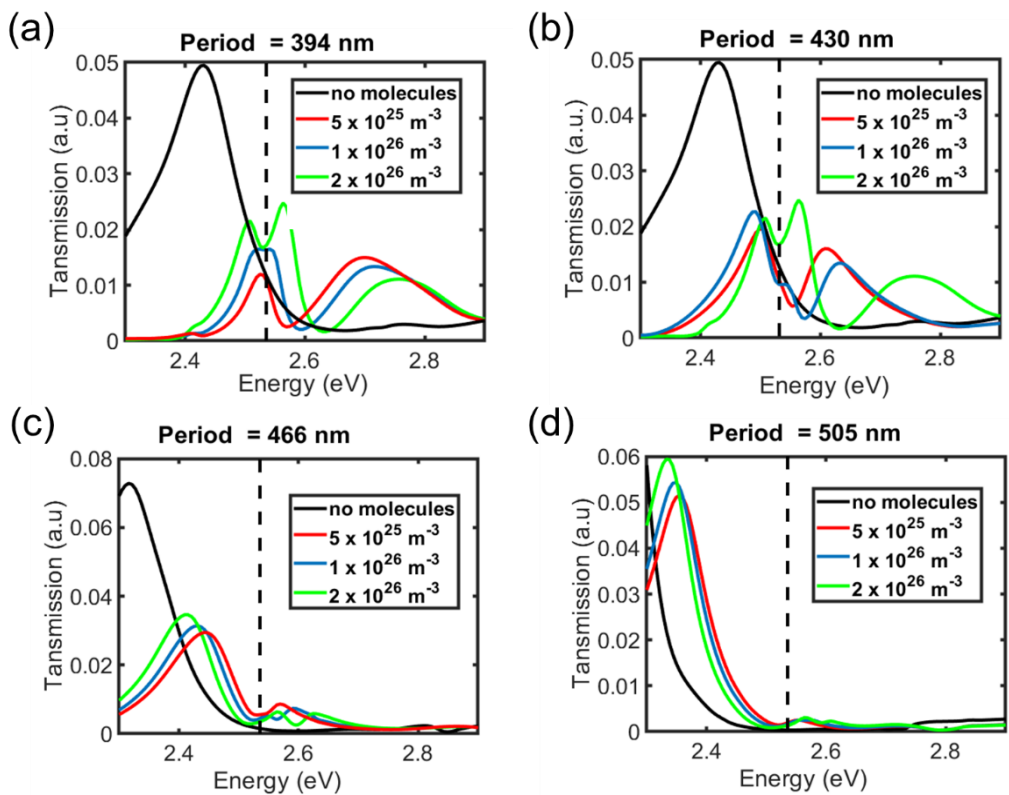

**Figure S3.** Simulations of the transmission spectra of the hybrid systems for 3 different molecular concentrations as is indicated in the insets, showing the appearance of the 3 peak at high concertation.

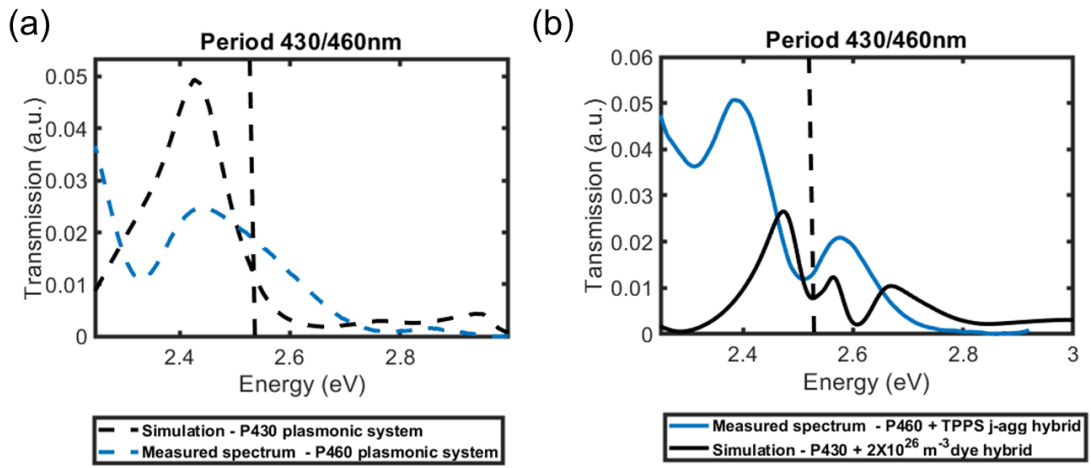

**Figure S4.** Comparison between the experimental and the simulation results in a plasmonic (a) and the hybrid system (b), showing a case in which the third peak does not appear in experiment, but only in the simulation.